%% file: main.tex
\documentclass[aps,prl,reprint,amssymb]{revtex4-2}

\usepackage{xcolor}

\usepackage{amsmath}
\usepackage{mathtools}
\usepackage{bm}

\usepackage{amsmath}  
\usepackage{amsfonts}
\usepackage{braket}
\usepackage{graphicx}
\usepackage{hyperref}
\hypersetup{
    colorlinks=true,
	citecolor=blue,
    linkcolor=red,
	urlcolor = blue
}

\begin{document}

\title{Engineering a driven-dissipative  bath of altermagnetic quantum magnons for controlling classical dynamics of spins hosting  spin waves, domain walls, or skyrmions}

\author{Felipe Reyes-Osorio}
\email{freyes@udel.edu}
\affiliation{Department of Physics and Astronomy, University of Delaware, Newark, DE 19716, USA}
\author{Branislav K. Nikoli\'c}
\email{bnikolic@udel.edu}
\affiliation{Department of Physics and Astronomy, University of Delaware, Newark, DE 19716, USA}


\begin{abstract}
Using Schwinger-Keldysh field theory (SKFT), we engineer a dissipative and driven (i.e., out of equilibrium) bosonic bath acting on  {\em classical} localized 
spins  within a ferromagnetic insulator (FI) layer whose dynamics is governed  by the Landau-Lifshitz-Gilbert equation, as is usually assumed in spintronics and magnonics. The bosonic bath is comprised of {\em quantum} magnons within a layer of altermagnetic insulator (AMI) that is attached to a conventional FI layer, often one of the key ingredients within spintronic and magnonic  multilayers, so that interaction between slow classical (in the FI layer) and fast quantum (in the AMI layer) localized spins ensues. Such a bath, including its driving to produce a nonequilibrium distribution of altermagnetic magnons, generates a rich structure of the SKFT-derived extended LLG equation  for classical  spins within the FI layer. Our LLG equation contains two damping terms, both of which are spatially {\em nonlocal} and {\em anisotropic}, while one of them is also intrinsically \textit{non-Markovian}, i.e., nonlocal in time. We demonstrate how to exploit these terms for tuning spintronic and magnonic effects within the FI layer of AMI/FI bilayers that involve  spin wave or domain wall propagation, as well as skyrmion annihilation.   
\end{abstract}

\maketitle

{\em Introduction}---The magnetization dynamics of nonequilibrium magnets, as the key ingredient of spintronic and magnonic effects and device applications~\cite{Chumak2015,Flebus2024, Pirro2021}, is inevitably accompanied by dissipation. The magnetization dissipation is routinely captured by the phenomenological Gilbert damping term~\cite{Landau1935, Gilbert2004,Evans2014}, $\alpha_G \mathbf{S}_n \times \partial_t \mathbf S_{n}$ [see  Eq.~\eqref{eq:llg}], which is local in space and time and parametrized by a spatially homogenous constant $\alpha_G$. This parameter is typically extracted from experiments~\cite{Weindler2014} or computed as a material property via first-principles calculations~\cite{Ebert2011,Liu2014a} considering spin-orbit coupling and/or magnetic disorder scattering~\cite{Starikov2010}. The damping of the dynamics of spins $\mathbf{S}_{n}(t)$, which in the LLG approach are viewed as classical vectors of fixed length localized at the lattice sites of magnetic materials, is often an impediment in device applications, e.g., it limits propagation of spin waves (SW) and complexity of circuits in magnonics~\cite{Chumak2015,Flebus2024, Pirro2021}. As such, a variety of intuitively justified experimental schemes have been intensely pursued to counteract it~\cite{Breitbach2023,Merbouche2024}, such as by injecting spin angular momentum via spin current~\cite{Merbouche2024}. 

However, recent rigorous derivations of extended LLG equations, such as from open quantum system theory~\cite{Anders2022,GarciaGaitan2024,ReyesOsorio2026} or from Schwinger-Keldysh nonequilibrium quantum field theory (SKFT)~\cite{Bhattacharjee2012, ReyesOsorio2024a,ReyesOsorio2025,Verstraten2023,Quarenta2024}, demonstrate the possibility of a much richer structure of magnetization damping that sensitively depends on the microscopic details of the dissipative bath acting on localized spins $\mathbf{S}_{n}$. For example, damping can be nonlocal in space or time, where the latter nonlocality, i.e., non-Markovianity of the extended LLG equation, harbors magnetic inertia~\cite{Bhattacharjee2012,Mondal2023,Neeraj2020,Bajpai2019,Quarenta2024,ReyesOsorio2025}, $\propto \mathbf{S}_n \times \partial^2_t \mathbf S_{n}$, depending on the second derivative of $\mathbf{S}_n(t)$~\footnote{Note that we use shorthand notation $\partial_t \equiv \partial/\partial_t$ and $\partial^2_t \equiv \partial^2/\partial_t^2$.}. In turn, magnetic inertia opens the possibility of new effects, such as  excitation of the so-called inertial SWs~\cite{ReyesOsorio2025,Mondal2022,dAquino2023} that are outside the band of SWs that can be excited thermally. 

\begin{figure}
    \centering
    \includegraphics[width=\columnwidth]{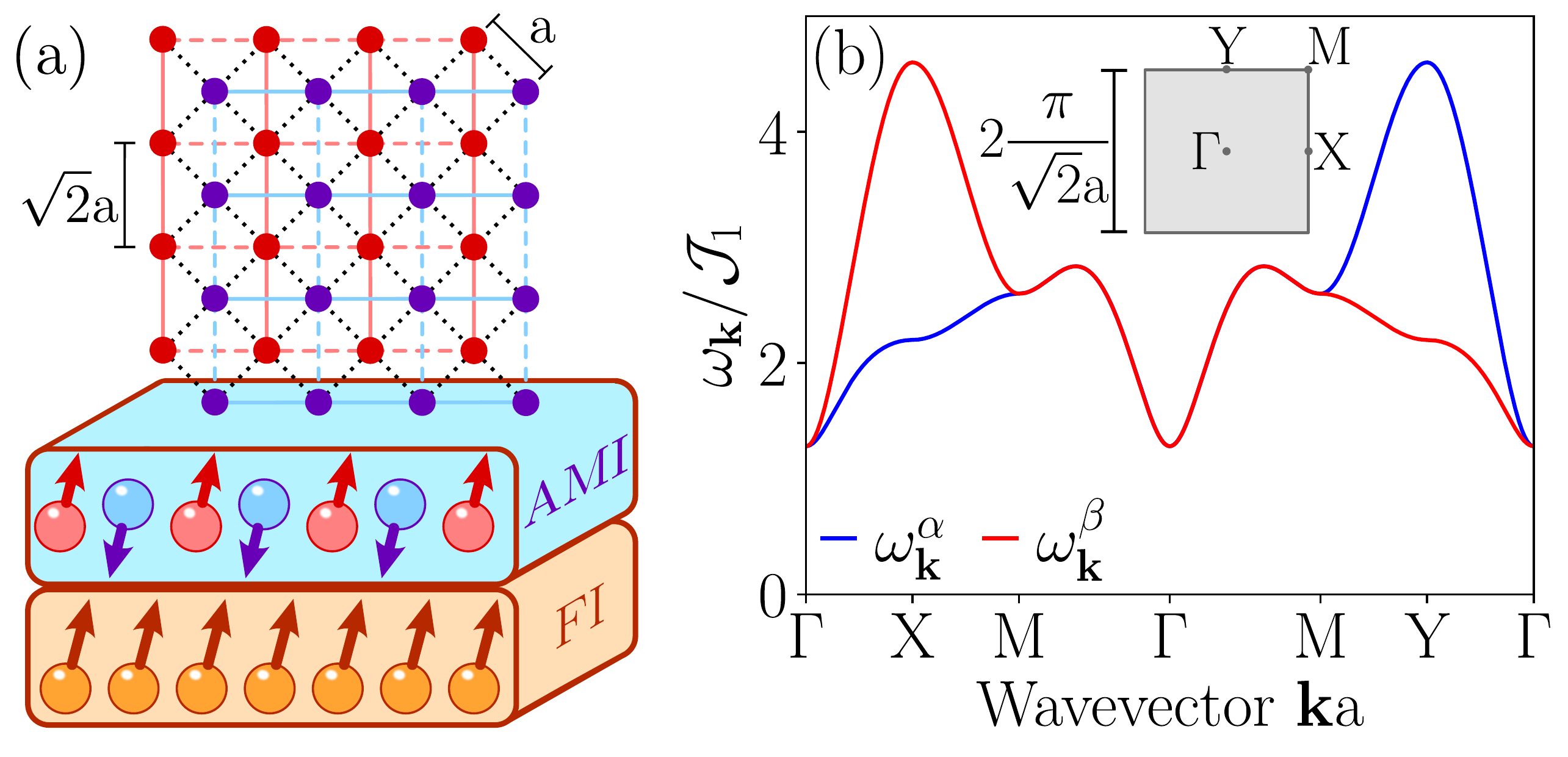}
    \caption{(a) Schematic view of the AMI/FI  bilayer, where the AMI layer hosts a bath of altermagnetic \textit{quantum} magnons arising  as the low-energy excitations of the Heisenberg Hamiltonian. The exchange coupling between NN localized spins (dotted black lines) is $\mathcal J_1>0$, while the one between NNN spins is $\mathcal J_2(1+\delta)$ (solid lines) or $\mathcal J_2(1-\delta)$ (dashed lines) depending~\cite{GarciaGaitan2025, Liu2025} on the direction and sublattice (blue for sublattice $A$ and red for sublattice $B$). (b) Spin-split spectrum of chiral altermagnetic magnons~\cite{Smejkal2023} computed for $\mathcal J_2=0.2\mathcal J_1$, $\delta=0.3$, and magnetic anisotropy $\mathcal K=0.1 \mathcal J_1$. The inset of panel (b) labels high-symmetry points in the first BZ.}
    \label{fig:setup}
\end{figure}

The possibility of engineering dissipative baths as a resource has been vigorously explored in quantum information science~\cite{Verstraete2009,Harrington2022}, but it remains underutilized~\cite{Breitbach2023, Takei2019} in spintronics and magnonics. In this Letter, we employ  SKFT~\cite{Kamenev2023,Altland2023,Gelis2019} to demonstrate how such newly acquired understanding of microscopic origins of magnetization damping can  open avenues to engineer it by proper choice of the bath surrounding classical spins $\mathbf{S}_n$, as well as by its additional driving out of equilibrium. For this purpose, we consider a driven-dissipative bath composed of {\em quantum} magnons~\footnote{Note that we use precise terminology, as required for the bilayer in Fig.~\ref{fig:setup} where classical localized spins in the FI layer coexist with quantum localized spins within the AMI layer, where ``SW'' stands for excitations of classical localized spins described by the LLG equation~\cite{Moreels2026,Kim2010} and we reserve the term ``magnon'' for their quantized counterparts hosted by the AMI layer.} hosted by an altermagnetic insulator (AMI)~\cite{Wan2025} layer and show how it can dramatically change LLG dynamics of classical spins within a ferromagnetic insulator (FI) layer of the AMI/FI bilayer illustrated by Fig.~\ref{fig:setup}. Note that FIs like yttrium iron garnet (YIG) are central to applications~\cite{Serga2010} due to their low intrinsic damping $\alpha_G$~\cite{Bertelli2021}. Altermagnets~\cite{Smejkal2022, Smejkal2022a, Bai2024, Wan2025} are a recently identified class of magnetic materials with zero net magnetization, akin to antiferromagnets, but also a host of other properties akin to ferromagnets. Thus, they operate at higher frequencies than ferromagnets, but are more controllable than antiferromagnets. Furthermore, magnons in AMIs can exhibit intricate spatial anisotropy~\cite{Biniskos2025, Liu2024, Hoyer2025, Beida2025, Yu2024} and strong mutual interactions~\cite{GarciaGaitan2025, Wiedmann2025, Jin2025, Cichutek2025, Cichutek2025a, Eto2025}. SKFT is employed to integrate out the bosonic bath of quantum magnons within the AMI layer in Fig.~\ref{fig:setup}, which leads to an unusual and not previously encountered~\cite{Evans2014} LLG equation for classical spins in the FI layer as our central result
\begin{align}\label{eq:llg}
    \partial_t \mathbf S_n &= - \mathbf S_n \times \bigg[ \tilde{\mathbf B}_n - \alpha_G \partial_t \mathbf S_{n} \\
    & -\sum_{n'} \Lambda_{nn'} \partial_t  S^z_{n'} \mathbf{e}_z + \int\! dt' \, \sum_{n'} \eta_{nn'}(t,t') \mathbf S_{n'}(t')\bigg]. \nonumber
\end{align}
The first two terms on the right-hand side (RHS) of Eq.~\eqref{eq:llg}, whose magnitudes are specified by the effective magnetic field $\tilde{\mathbf{B}}_n$ and Gilbert damping parameter $\alpha_G$, are the conventional LLG equation~\cite{Landau1935, Gilbert2004, Evans2014} for classical spins $\mathbf{S}_n$, as widely implemented in micromagnetic~\cite{Moreels2026} and atomistic spin dynamics~\cite{Evans2014} codes. Then, the backaction of the surrounding AMI quantum magnons introduces terms that are {\em spatially nonlocal}, i.e., coupling distant spins at locations $n\neq n'$; and can be either Markovian (third term on RHS) or non-Markovian (fourth term on RHS), where the latter type of terms has also attracted very recent experimental attention~\cite{Hartmann2025}. In general, and unlike in some other SKFT-based derivations~\cite{Bhattacharjee2012,Verstraten2023, Quarenta2024, ReyesOsorio2024a}, the non-Markovian term \textit{cannot} be Taylor-series expanded and truncated to extract time-local damping or inertia. Instead, numerical solutions of Eq.~\eqref{eq:llg} must handle it explicitly~\cite{Scali2024}.

\begin{figure}
    \centering
    \includegraphics[width=\columnwidth]{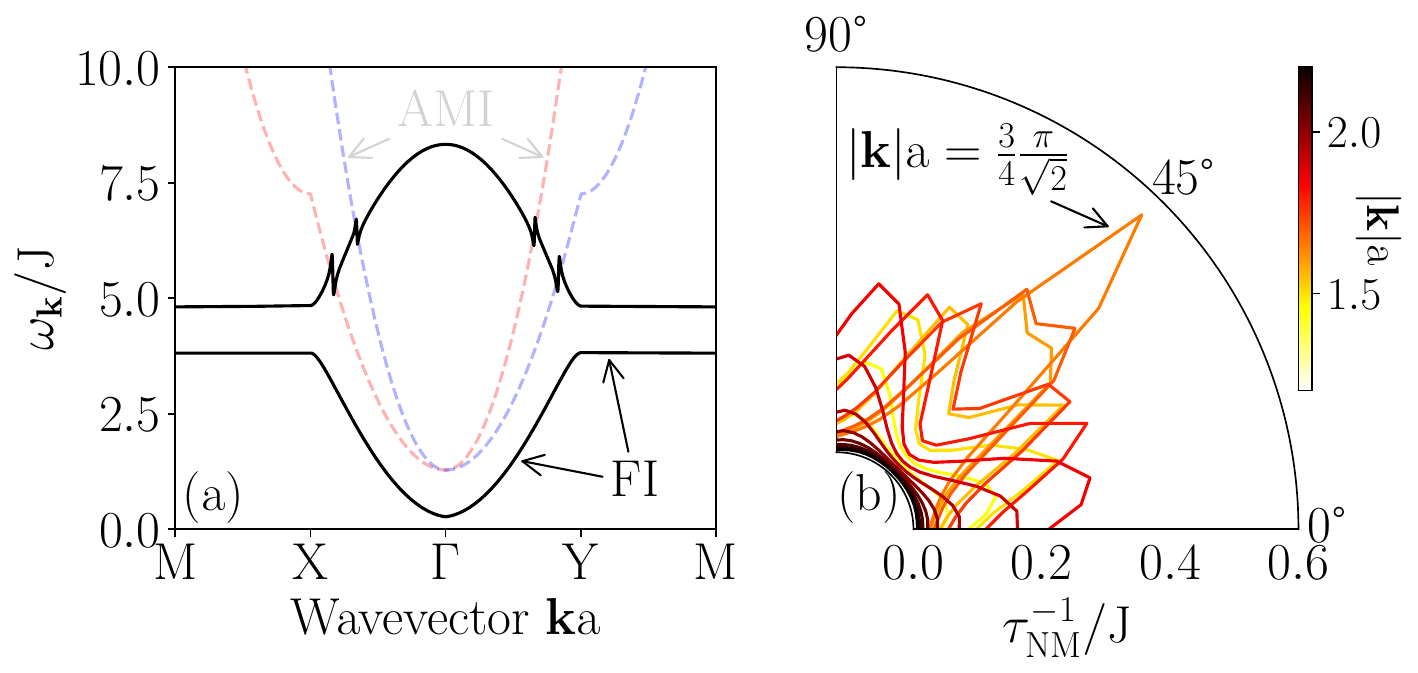}
    \caption{(a) SW spectrum of FI layer (black lines) when subjected to interaction with the bath of quantum magnons in AMI layer [Fig.~\ref{fig:setup}(a)]. At the crossing points, spikes emerge in the upper SW band, while SW damping increases (b) with intricate dependence on wavevector amplitude and direction. FI parameters are chosen as $J=\mathcal{J}_1$ and $K=-0.3 J$.}
    \label{fig:swfig}
\end{figure}

\begin{figure}
    \centering
    \includegraphics[width=\columnwidth]{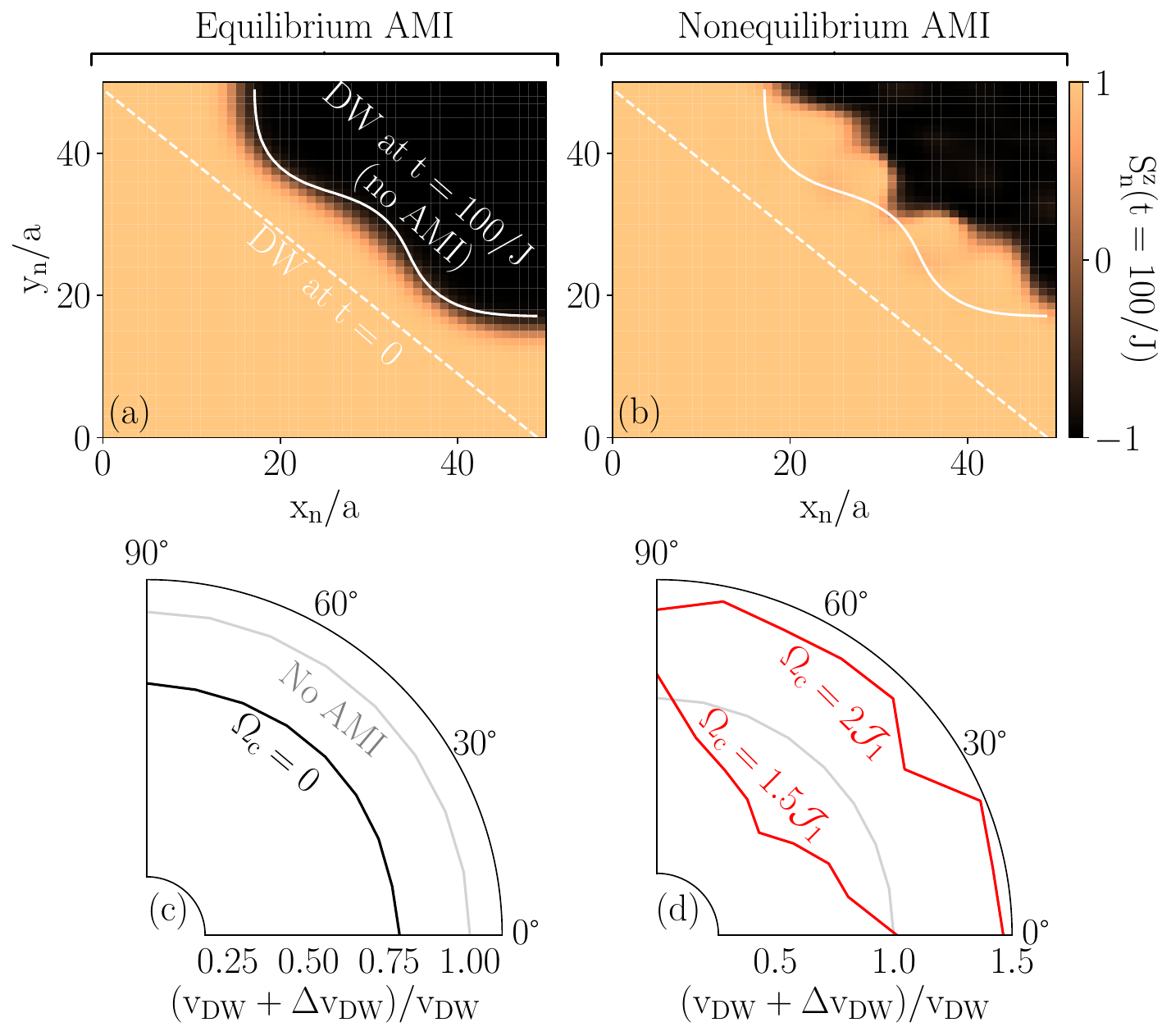}
    \caption{Spatial distribution of localized spins $S^z_n(t)$ within the FI layer at a time $t=100/J$ after its DW propagation subjected to the bath of AMI quantum magnons [Fig.~\ref{fig:setup}] that is in (a) thermal equilibrium; or (b) driven at frequency $\Omega_c=2\mathcal{J}_1$. Initial and final DW positions, in the absence of the AMI layer, are indicated in dashed and solid white lines, respectively. Panels (c) and (d) show the relative change of DW velocity as a function of its direction of propagation when AMI is in vs.  out of equilibrium, respectively.}
    \label{fig:dw}
\end{figure}

\begin{figure}
    \centering
    \includegraphics[width=\columnwidth]{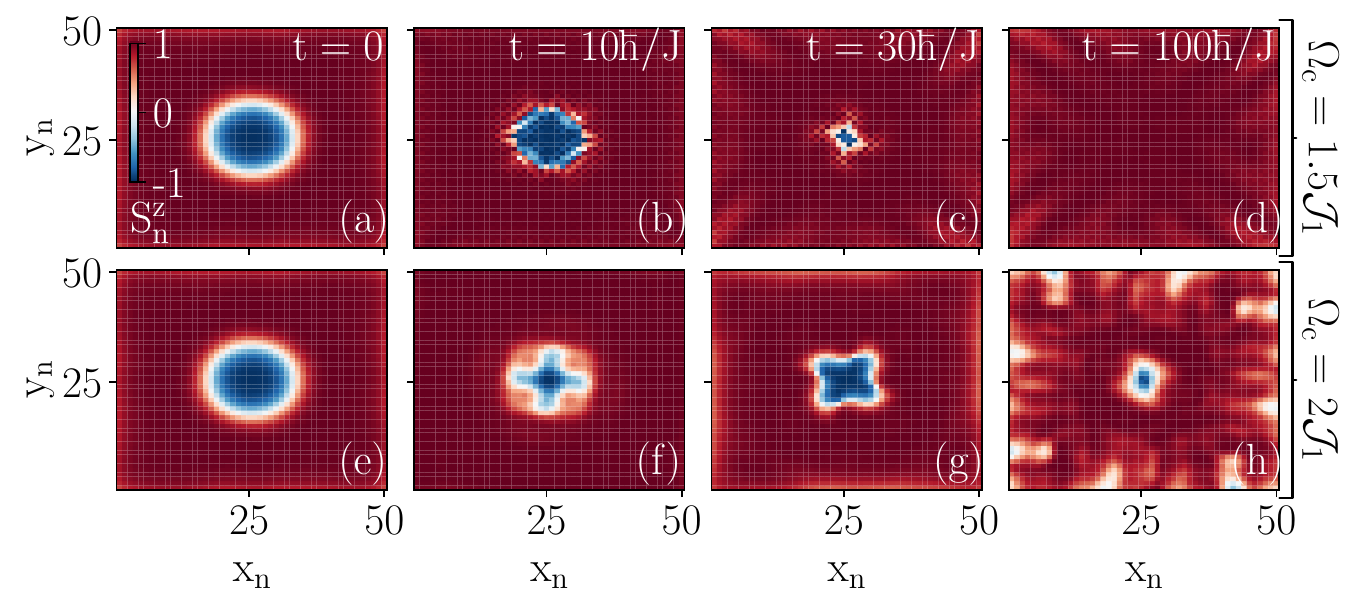}
    \caption{Spatial distribution of localized spins $S^z_n(t)$ at different times within the FI layer  subjected to the bath of AMI quantum magnons [Fig.~\ref{fig:setup}]. The bath is driven at frequency (a)--(d) $\Omega_c=1.5\mathcal{J}_1$; or (e)--(h) $\Omega_c=2\mathcal J_1$.} 
    \label{fig:skyrmion}
\end{figure}

We emphasize that our SKFT-based approach to a many-body systems  mixing quantum (as they could be hosted by a layer of antiferromagnet~\cite{Scheie2021,Yamaguchi2024,GarciaGaitan2024,Koerber2025,Lenzing2023}, altermagnet~\cite{Cichutek2025,GarciaGaitan2025} or more exotic magnets like quantum spin liquid~\cite{Grover2013,Lyu2025,GarciaGaitan2026}) and classical (as hosted by conventional ferromagnetic layers like YIG~\cite{GarciaGaitan2024,Quarenta2024,Serga2010}) localized spins also fills a gap in the literature where there has been considerable experimental interest in such systems~\cite{Yamaguchi2024,Gonzalez-Ballestero2022}, but with no general theoretical framework~\cite{Lenzing2023} being developed thus far~\footnote{Note that flowing spins of conduction electrons are always fast and must be treated quantum mechanically~\cite{Bajpai2019}.}. Our extended LLG equation~\eqref{eq:llg} applied to spins within the FI layer in Fig.~\ref{fig:setup}, predicts how SWs [Fig.~\ref{fig:swfig}], magnetic domain  walls (DWs) [Fig.~\ref{fig:dw}] or skyrmions [Fig.~\ref{fig:skyrmion}] carried by them can be controlled by tailoring unconventional third and fourth terms on its RHS. Before explaining these predictions, we first introduce useful concepts and notation.

\textit{Models and Methods}---We model localized spins within the AMI/FI bilayer in Fig.~\ref{fig:setup} using a Hamiltonian \mbox{$\hat H = \hat H_{\rm FI} + \hat H_{\rm AMI} + J_I\sum_n \hat{\mathbf S}_n \cdot \hat{\mathbf{s}}_n$}. Here
\begin{equation}\label{eq:afspinham}
    \hat{\mathcal H}_{\rm AMI} = \sum_n \mathcal{J}_{nn'} \hat{\mathbf{s}}_n \cdot \hat{\mathbf{s}}_{n'} + \mathcal{K}\sum_n \left(\hat s^z_n \right)^2,
\end{equation}
is the Hamiltonian of quantum spins within the AMI layer; $\hat{\mathbf{s}}_n$ is the spin operator at the site $n$ of a two-dimensional (2D) Lieb lattice~\cite{Liu2025,GarciaGaitan2025}; $\mathcal K<0$ is the easy-axis magnetic anisotropy;  and $\mathcal{J}_{nn'}$ is the exchange coupling, which is $\mathcal{J}_{nn'}=\mathcal J_1>0$ between the nearest-neighbors (NN) and  $\mathcal{J}_{nn'}=\mathcal J_2(1\pm\delta)/2$ between the next-nearest-neighbors (NNN),  depending on direction and sublattice [Fig.~\ref{fig:setup}(a)]. The ground state of this model exhibits N\'eel order along the $z$ direction and becomes altermagnetic when $\delta \neq 0$~\cite{Liu2025}. To describe its low-energy dynamics, we map the spin operators to bosonic ones via the standard linearized Holstein-Primakoff transformation~\cite{Bajpai2021, Gohlke2023, GarciaGaitan2025},  
$\hat{s}^z_{n \in A} = s - \hat a_n^\dagger \hat a_n$;  $\hat s^-_{n \in A} = \sqrt{2s}\hat a_n$; 
$\hat s^z_{n \in B} = -s + \hat b_n^\dagger \hat b_n$; and  $\hat s^-_{n \in B} = \sqrt{2s}\hat b_n^\dagger$; where $\hat a_n$ ($\hat b_n$) are bosonic operators on the $A$ ($B$) sublattice, respectively. The linearized Holstein-Primakoff transformation becomes exact in the large spin value $s$ limit~\cite{Bajpai2021}. The resulting bosonic Hamiltonian is quadratic, so it can be diagonalized via the Bogoliubov transformation~\cite{Xiao2009}, becoming
$\hat{H}_{\rm AMI} = s\sum_\mathbf{k} \left(\omega^\alpha_\mathbf{k} \hat\alpha_\mathbf{k}^\dagger\hat\alpha_\mathbf{k} + \omega^\beta_\mathbf{k} \hat\beta_\mathbf{k}^\dagger\hat\beta_\mathbf{k} \right)$, 
where $\hat \alpha_\mathbf{k}$ and $\hat \beta_\mathbf{k}$ are magnon operators of opposite chiralities~\cite{Smejkal2023} with wavevector $\mathbf{k}$ carrying angular momentum $\pm \hbar$, respectively. Their dispersions $\omega^{\alpha,\beta}_\mathbf{k}$ [Fig.~\ref{fig:setup}(b)] exhibit $d$-wave splitting with nodes along the $\Gamma$--$M$ direction in the first Brillouin zone (BZ)~\cite{GarciaGaitan2025}. The term $\hat H_{\rm FI}$ describes localized spins $\hat{\mathbf{S}}_n$ (initially considered as quantum) within the FI layer, and $J_I$ is the interlayer exchange coupling.

From the Hamiltonian of the bilayer system, we construct the SKFT action \mbox{$S=S_{\rm FI} + S_{\rm AMI}$}, where $S_{\rm FI}$ is the action of a 2D FI in the limit of small quantum fluctuations~\cite{Altland2023}
\begin{equation}
    S_{\rm FI} = \int\! dt \, \sum_n \mathbf S_{n}^q\cdot \left(\partial_t \mathbf S_{n}^c \times \mathbf{S}_{n}^c + \mathbf{B}^{\rm eff}_n[\mathbf{S}^c_{n}]\right). \label{eq:actionFM}
\end{equation}
Here, $\mathbf{S}^{c,q}_n$ are real vector fields corresponding to the spin expectation value of spin coherent states~\cite{Altland2023} and the superscripts $c,q$ denote the classical and quantum components, respectively, obtained from linear combinations of the fields evaluated in different segments of the Keldysh closed time contour~\cite{Kamenev2023}. Additionally, the first term on the RHS of Eq.~\eqref{eq:actionFM} stems from the Berry phase~\cite{Altland2023}, where $\mathbf{B}^{\rm eff}_n[\mathbf{S}_n^c]=-\delta H_{\rm FI}/\delta \mathbf{S}_n \big|_{\mathbf{S}_n^c}$ is the effective magnetic field defined by the classical Hamiltonian of localized spins within the FI layer 
\begin{equation}\label{eq:fiham}
    H_{\rm FI} = -J \sum_{\braket{nn'}}\mathbf{S}_n \cdot \mathbf{S}_{n'} + K \sum_n (S^z_n)^2.
\end{equation}
Here, $J>0$ and $K<0$ are the ferromagnetic exchange coupling and easy-axis magnetic anisotropy in the FI, respectively, and $\braket{nn'}$ indicates summation over NN. On the other hand, $S_{\rm AMI}$ [Eq.~\eqref{eq:actionrotated} in End Matter] is the action of the AMI layer including the interlayer exchange interaction, which contains terms that are at most quadratic in the bosonic magnon operators. Thus, $S_{\rm AMI}$ can be integrated out from the SKFT functional integral to second order in $J_I$, yielding an effective action $S_{\rm eff}$ [Eq.~\eqref{eq:effaction} in End Matter] for the FI fields only subjected to the backaction of AMI magnons. 

The extended LLG equations are extracted~\cite{Bhattacharjee2012,Verstraten2023, Quarenta2024, ReyesOsorio2024a} from $S_{\rm eff}$ via the saddle point approximation \mbox{$\delta S_{\rm eff}/\delta \mathbf{S}^q=0$}, which can also be justified diagrammatically~\cite{ReyesOsorio2026}. Within the extended LLG equation~\eqref{eq:llg}, \mbox{$\tilde{\mathbf{B}}_n=\mathbf{B}^{\rm eff}_n + B^{\rm stag}_n \mathbf{e}_z$} contains the effective magnetic field of the FI and a staggered magnetic field $B^{\rm stag}\mathbf{e}_z$ induced by the AMI,  whose sign alternates on different sublattices; $\Lambda_{nn'}$ is a nonlocal function coupling distant spins via a field-like torque \mbox{$\propto \partial_t S^z_{n'}$}; and the non-Markovian memory tensor is given by
\begin{equation}\label{eq:memorytensor}
    \eta_{nn'}(t-t')=\begin{pmatrix}
    \eta_{nn'}^\|(t-t') & \eta_{nn'}^\perp(t-t') & 0  \\
    -\eta_{nn'}^\perp(t-t') & \eta_{nn'}^\|(t-t') & 0 \\
    0 & 0 & 0
    \end{pmatrix}.
\end{equation}
In terms of the AMI spin length $s$ and interlayer coupling $J_I$, which are the expansion parameters of the theory, $B^{\rm stag}\propto sJ_I$, $\eta_{nn'}(t-t') \propto sJ_I^2$ and $\Lambda_{nn'} \propto J_I^2$. Physically, the staggered field is induced by the AMI N\'eel order, and its strength is decreased by thermal occupation of magnons. The non-Markovian torque containing the memory tensor $\eta_{nn'}(t-t')$ stems from direct magnon exchange processes between the layers, and it is non-Markovian because such processes involve excitations at similar frequencies. On the other hand, the nonlocal field-like torque $\propto \Lambda_{nn'}$ originates from density-density interactions between the layers, leading to a Markovian term since AMI magnons are much faster than low-energy  excitations of a typical FI. Due to the nature of the processes generating terms in our extended LLG equation~\eqref{eq:llg}, $\eta_{nn'}(t-t')$ depends only on the AMI spectrum via the retarded and advanced AMI Green's functions (GFs), whereas $B^{\rm stag}_n$ and $\Lambda_{nn'}$ in addition require the AMI magnon distribution function $\mathcal F(\omega)$ via the Keldysh GF~\cite{Kamenev2023, Altland2023, Gelis2019} (see End Matter for their explicit expressions). For simplicity, we assume that the same $\mathcal F(\omega)$ applies to both AMI magnon chiralities, which in thermal equilibrium at temperature $T$ becomes \mbox{$\mathcal F_{\alpha,\beta}(\omega)=\coth\left(\omega/2T\right)$}~\footnote{Note that we use $\hbar=k_B=1$ for simplicity.}. In addition, we model the excitation of nonequilibrium magnons at a frequency $\Omega_c$ within AMI, as achieved by drive via, e.g., a microwave antenna~\cite{Chumak2015}, phenomenologically by using $\mathcal F(\omega) \propto \epsilon/[(\omega-\Omega_c)^2 + \epsilon^2]$, where $\epsilon\ll\Omega_c$ indicates coherent excitation~\cite{Pirro2021}.

\begin{figure}
    \centering
    \includegraphics[width=\columnwidth]{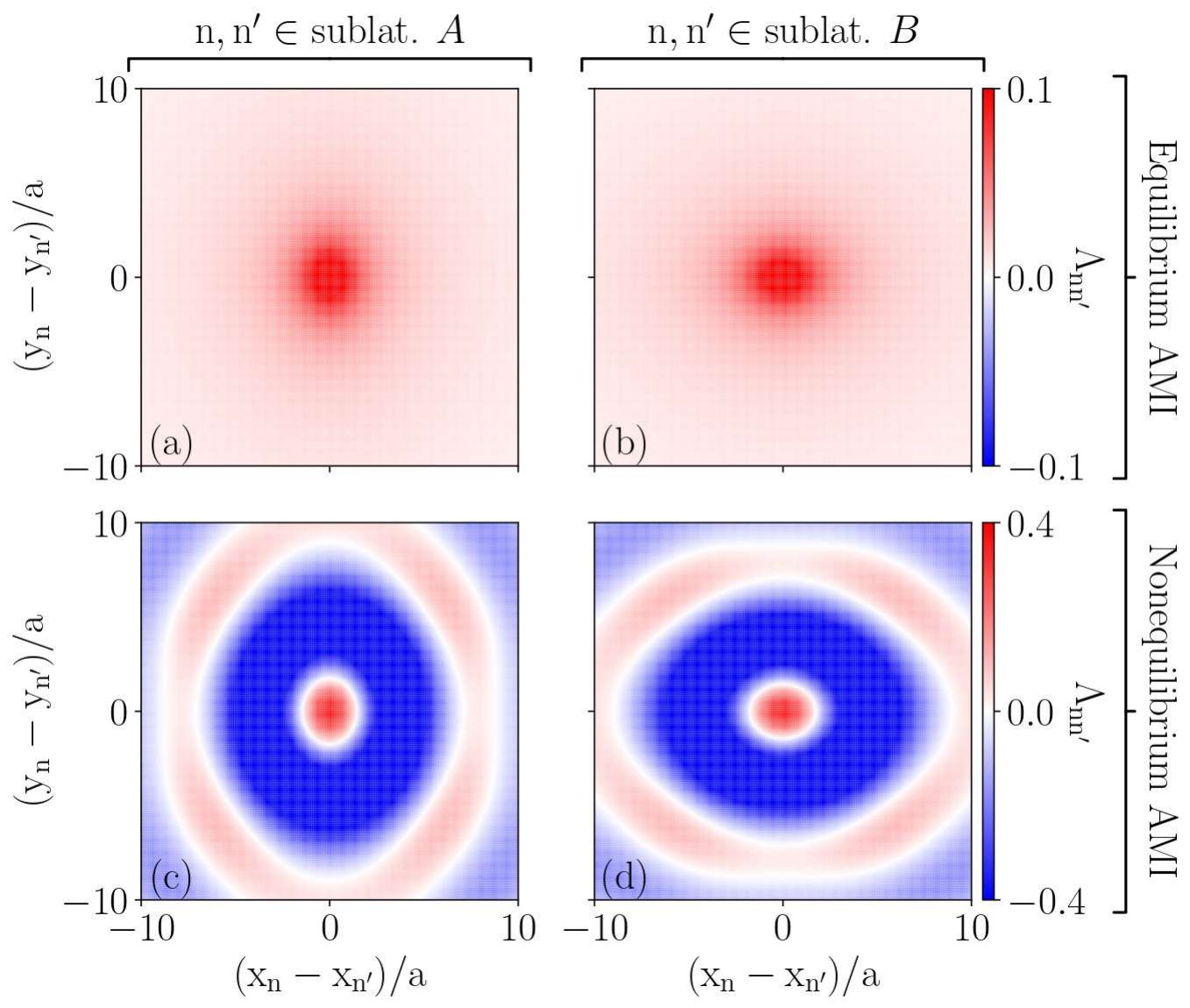}
    \caption{Spatial dependence of nonlocal function $\Lambda_{nn'}$ in Eq.~\eqref{eq:llg}, as induced by the bath of AMI quantum magnons in (a),(b) equilibrium at temperature, $T=\mathcal J_1$; or (c),(d) out of equilibrium due to driving coherent magnons at frequency $\Omega_c = 2\mathcal J_1$. Darker colors indicate stronger coupling of spins at relative separation $\mathbf r = \mathbf r_n - \mathbf r_{n'}$. Panels in the left (right) column depict coupling within sublattice $A$ ($B$), while inter-sublattice coupling is not shown. AMI parameters are chosen as  $\mathcal J_2 = 0.2\mathcal J_1$, $\delta = 0.3$ and $\mathcal{K}=-0.1\mathcal{J}_1$.}
    \label{fig:lambda}
\end{figure}

\textit{Results and Discussion}---To showcase the potential of AMI for dissipation engineering, we first examine the SW spectrum of the FI layer in Fig.~\ref{fig:setup} by linearizing Eq.~\eqref{eq:llg}. Because of space and time translation invariance, SWs decouple in wavevector and frequency space even when accounting for the non-Markovian tensor, and we solve the resulting eigenvalue problem numerically. The resulting SW spectrum is plotted in Fig.~\ref{fig:swfig}(a). Because of the two-sublattice structure imprinted by the AMI layer on the FI layer, the BZ is reduced,  and a gap appears due to the staggered magnetic field $B^{\rm stag}_n\mathbf{e}_z$. In addition, the SW spectrum is further modified [spikes in Fig.~\ref{fig:swfig}(a)] at crossing points with the AMI spectrum [light red and blue curves in Fig.~\ref{fig:swfig}(a)]. Crossings between the spectra are only possible in the upper SW band, where spikes indicate hybridization with the AMI magnons, which would appear as avoided crossings in a theory that does not integrate out the environment. Note that we are considering a long wavelength approximation for the AMI magnon dispersion $\omega^{\alpha,\beta}_\mathbf{k}$ in Fig.~\ref{fig:setup}(b) to be able to make progress analytically. The SW damping $\tau^{-1}$, as extracted from the imaginary part of the FI SW frequencies, increases at the crossing points since the SW is able to transfer energy to the magnon bath. The change in SW damping---\mbox{$\tau^{-1}_{\rm NM} = \tau^{-1}-\tau^{-1}_G $}, where $\tau^{-1}_G$ is due only to Gilbert damping---due to the non-Markovian AMI-induced torque  is plotted in Fig.~\ref{fig:swfig}(b). We find that $\tau^{-1}_{\rm NM}$ depends intricately on the direction and magnitude of the wavevectors at which the spectra cross, reaching a maximum when $\mathbf{k}$ has magnitude $3\pi/4\sqrt 2 a$ and points along an AMI nodal direction.

Although the field-like nonlocal torque induced by the AMI layer is not strictly Markovian at high SW frequencies at which hybridization occurs, the whole term can nonetheless be neglected due to being higher order in $1/s$. However,  Fig.~\ref{fig:swfig} shows that the memory tensor $\eta_{nn'}(t-t')$ is not operative for low-energy FI excitations, at which point the effects of $\Lambda_{nn'}$ dominate and are indeed Markovian. We plot the nonlocal function $\Lambda_{nn'}$ vs. the relative position vector $\mathbf r_n - \mathbf r_{n'}$ in Fig.~\ref{fig:lambda}. When the AMI magnons are in thermal equilibrium [Figs.~\ref{fig:lambda}(a) and~\ref{fig:lambda}(b)], the nonlocality is short-ranged, coupling spins separated by \mbox{$\lesssim$ 5} sites. However, nonlocality drastically increases when driving the AMI layer to produce nonequilibrium magnons at frequency $\Omega_c$ [Figs.~\ref{fig:lambda}(c) and~\ref{fig:lambda}(d)]. In and out of equilibrium, $\Lambda_{nn'}$ elongates along alternating directions for $n,n'$ both in sublattice $A$ or $B$, reflecting the $d$-wave AMI splitting. When $n,n'$ are in different sublattices, $\Lambda_{nn'}$ is isotropic (not shown).

While the nonlocal field-like torque has no appreciable effect on SWs, it dramatically modifies the velocity of DWs in Fig.~\ref{fig:dw}, which is of great interest to applications~\cite{Catalan2012, Kim2017d}. To highlight this modification, we simulate the propagation of DWs in a $50 \times 50$-site lattice. Since DWs are low-energy states of the ferromagnet, we neglect the non-Markovian torque in Eq.~\eqref{eq:llg}. For simplicity, we drive the DWs with an external field, so that they move at velocity $v_{\rm DW}$ in the absence of the AMI layer. When the AMI layer in equilibrium is added, the nonlocal field-like torque slows down DW propagation [Fig.~\ref{fig:dw}(a)]. On the other hand, $v_{\rm DW}$ increases for driven AMI layer hosting nonequilibrium magnons at frequency $\Omega_c = 2 \mathcal J_1$ [Fig.~\ref{fig:dw}(b)]. We use the total change in magnetization as a measure of the relative change in DW velocity [Figs.~\ref{fig:dw}(c) and~\ref{fig:dw}(d)]
\begin{equation}
    \frac{v_{\rm DW} + \Delta v_{\rm DW}}{v_{\rm DW}} = \frac{\sum_n \left[S^z_n(t=100/J) - S^z_n(0)\right]}{\sum_n \left[S^z_n(t=100/J) - S^z_n(0)\right]_{\textrm{No AMI}}}.
\end{equation}
When driving the AMI at frequency $\Omega_c=1.5\mathcal J_1$, the extent of the nonlocality of $\Lambda_{nn'}$ is commensurable with the width of the DWs we employ in our simulations. Therefore, at this frequency, the DW can resolve the spatial anisotropy of $\Lambda_{nn'}$, resulting in DW velocity that is slowed down drastically along the AMI nodal directions [Fig.~\ref{fig:dw}(d)].

Engineering dissipation of the bosonic bath of AMI quantum magnons can also manipulate other magnetic solitons, such as skyrmions~\cite{Fert2013, Nagaosa2013, Koshibae2015} whose propagation through the FI layer alone has been studied via SKFT before~\cite{Loss2017}. In our AMI/FI bilayer, skyrmions are stable when subjected to interaction with the bath of thermal AMI magnons. However, when we employ a driven AMI layer hosting nonequilibrium magnons, we find skyrmion \textit{annihilation} in Fig.~\ref{fig:skyrmion} which proceeds faster for frequency $\Omega_c=1.5\mathcal{J}_1$ than $\Omega_c=2\mathcal{J}_1$. At first sight, this is  surprising, considering that skyrmions are swirling textures of the local magnetization with topological protection against their unwinding~\cite{Fert2013, Nagaosa2013}. However, such protection has been typically confirmed using the standard LLG equation simulating skyrmion interaction with static disorder~\cite{Fert2013, Koshibae2015} rather than via many-body interactions we consider. Driving the AMI layer at frequency $\Omega_c=1.5\mathcal{J}_1$ annihilates the skyrmion faster than for $\Omega_c=2\mathcal{J}_1$. In addition, the nonlocality of the damping function $\Lambda_{nn'}$ [Fig.~\ref{fig:lambda}], increasing in range when the AMI layer is driven at frequency $\Omega_c=2\mathcal{J}_1$, leads to the skyrmion developing transient lobes [Figs.~\ref{fig:skyrmion}(f) and~\ref{fig:skyrmion}(g)] reminiscent of the $d$-wave nature of the overlayer. We note that skyrmion annihilation is one of the  crucial operations required in their memory applications~\cite{Koshibae2015}, so that the mechanism unraveled by Fig.~\ref{fig:skyrmion} offers an unforeseen route for that purpose.

\textit{Conclusions and Outlook}---The SKFT-derived highly unusual LLG equation~\eqref{eq:llg} suggests the  transformative potential of engineering of dissipative baths in spintronics and magnonics. Unlike the demands on such engineering in, e.g., quantum information science~\cite{Verstraete2009,Harrington2022}, in spintronics and magnonics, bath engineering can be achieved as simply as by introducing a proper magnetic layer and driving it out of equilibrium, as  suggested by Fig.~\ref{fig:setup}. When the  layer hosting a bath is chosen as AMI, the dynamics of SWs, DWs, and skyrmions in the FI layer of the AMI/FI bilayer in Fig.~\ref{fig:setup} can be dramatically modified [Figs.~\ref{fig:swfig},~\ref{fig:dw} and~\ref{fig:skyrmion}]. The nonlocal function $\Lambda_{nn'}$ [Fig.~\ref{fig:lambda}]  characterizing the field-like torque  encodes information about both the nonequilibrium drive and the AMI spin splitting, and it could be directly measured via relaxometry~\cite{Mzyk2022,Rovny2024} using a pair~\cite{Rovny2022} of spin qubits initially aligned in the plane of the bilayer. We anticipate that further knobs for controlling SWs in the FI layer could be activated by different types of driving of the bosonic bath in the AMI layer such that its magnon spectrum is modified, in addition to magnon occupation that we modified in this study. For example, periodic driving of the AMI layer may allow for amplification~\cite{Breitbach2023}, or even full suppression~\cite{Gonzalez-Ballestero2022}, of low-energy SWs hybridizing with the resulting AMI Floquet bands. We relegate such explorations to future studies.

This work was supported by the US Department of Energy (DOE) Grant No. DE-SC0026068.

\input{main.bbl}

\section{End Matter}

\subsection{SKFT of the AMI/FI bilayer system}

The spin Hamiltonian of Eq.~\eqref{eq:afspinham} in the bosonic momentum representation becomes
\begin{equation}\label{eq:hpham}
    \hat{\mathcal H}_{\rm AMI}= s\sum_k \hat{\phi}^\dagger_k \mathcal M_k \hat{\phi}_k,
\end{equation}
where $\sum_\mathbf{k}$ is over the first BZ, $\phi_k=(\hat a_k, \hat b_k, \hat a^\dagger_{-k}, \hat b^\dagger_{-k})^T$ is the bosonic Nambu spinor operator~\cite{Xiao2009, Okuma2022}, and $\mathcal M_k$ is a $4\times 4$ matrix
\begin{equation}
    \mathcal M_k = \begin{pmatrix}
        \gamma_k^A & 0 & 0 & \bar\gamma^{AB}_k \\
        0 & \gamma_k^B & \bar\gamma^{AB}_{-k} & 0 \\
        0 & \gamma^{AB}_{-k} & \gamma_k^A  & 0 \\
        \gamma^{AB}_k & 0 & 0 & \gamma_k^B 
    \end{pmatrix}.
\end{equation}
whose matrix elements are
\begin{subequations}\label{eq:gammas}
\begin{align}
    \gamma^{A,B}_k &= 2(\mathcal J_1-\mathcal J_2) + \mathcal J_2 \left[\cos\left(\sqrt 2 k_x\right) + \cos\left(\sqrt 2 k_y\right)\right] \nonumber \\
    &\mp 2\delta \left[\cos\left(\sqrt 2 k_x\right) - \cos\left(\sqrt 2 k_x\right)\right] - \mathcal{K}, \label{eq:gammaintra} \\
    \gamma^{AB}_k &= 2\mathcal J_1 e^{\frac{i}{\sqrt 2}(k_x+k_y)}\cos\left( \frac{k_x}{\sqrt 2}\right) \cos\left(\frac{k_y}{\sqrt 2} \right). \label{eq:gammainter}
\end{align}
\end{subequations}
The dispersions of the chiral magnons $\omega^{\alpha,\beta}_\mathbf{k}=\Delta_\mathbf{k} \mp (\gamma^A_\mathbf{k} - \gamma^B_\mathbf{k})$, where \mbox{$\Delta_k=\sqrt{(\gamma^A_k + \gamma^B_k)^2 - |\gamma^{AB}_k|^2}$}, are obtained by applying a paraunitary transformation $\mathcal T_k$~\cite{Xiao2009, Zhang2022} to $\mathcal M_\mathbf{k}$, such that $\mathcal{D}_\mathbf{k} = \mathcal{T}_\mathbf{k}^\dagger \mathcal{M}_\mathbf{k} \mathcal{T}_\mathbf{k}$ is diagonal.

We construct the SKFT based on such not-fully-diagonal form of Hamiltonian $\hat{\mathcal{H}}_{\rm AMI}$ in Eq.~\eqref{eq:hpham} to make the ensuing perturbation theory more tractable. Therefore, the SKFT action is
\begin{equation}\label{eq:actioncontour}
    S_{\rm AFI}^{(0)} = \int_\mathcal{C} \! dt \, \sum_\mathbf{k} \phi^\dagger_\mathbf{k} \left( \frac{i}{2}\tilde I \partial_t - \mathcal M_\mathbf{k} \right) \phi_\mathbf{k},
\end{equation}
where $\mathcal{C}$ is the closed Schwinger-Keldysh contour~\cite{Kamenev2023}, and $\phi_\mathbf{k}$ is a complex four-component field obtained as the eigenvalue of the bosonic coherent states $\hat \phi_\mathbf{k}\ket{\dots\phi_\mathbf{k}\dots} = \phi_\mathbf{k}\ket{\dots\phi_\mathbf{k}\dots}$. Contour time is replaced by real time via the Keldysh rotation~\cite{Kamenev2023} whereby the fields in the forward ($+$) and backward ($-$) segments of $\mathcal{C}$ become $\phi_\mathbf{k}^\pm = (\phi_\mathbf{k}^c \pm \phi_\mathbf{k}^q)/\sqrt 2$. The action then becomes
\begin{equation}\label{eq:actionrotated}
    S_{\rm AMI}^{(0)} = \int\! dtdt' \, \sum_\mathbf{k} \boldsymbol{\phi}^\dagger_\mathbf{k}(t) \check G_\mathbf{k}^{-1}(t,t') \boldsymbol{\phi}_\mathbf{k}(t'),
\end{equation}
where vectors in Keldysh space are denoted by bolded symbols, such as $\boldsymbol{\phi}_\mathbf{k} = (\phi_\mathbf{k}^c, \phi_\mathbf{k}^q)^T$, and matrices in Keldysh space are denoted by a check, such as the inverse GF, which in the energy representation is given by
\begin{equation}\label{eq:inverseGF}
    \check G^{-1}_\mathbf{k}(\omega) = \begin{pmatrix}
        0 & (\omega - i\eta)\frac{\tilde I}{2} - \mathcal M_\mathbf{k} \\
        (\omega + i\eta)\frac{\tilde I}{2} - \mathcal M_\mathbf{k} & i\eta \tilde{\mathcal F}(\omega)
    \end{pmatrix}.
\end{equation}
Here, $\tilde I=\textrm{diag}(1,1,-1-1)$ is the paraunitary identity matrix; $\eta$ is a positive infinitesimal; the $cq$ and $qc$ elements are the inverse advanced $(G^A_\mathbf{k})^{-1}$ and retarded $(G^R_\mathbf{k})^{-1}$ GFs, respectively, encoding the magnon spectrum; and the $qq$ element encodes the magnon distribution. Inverting $\check G_\mathbf{k}$ is done via transformation $\mathcal T_\mathbf{k}$, so that the AMI magnon GFs are given by
\begin{subequations}
\begin{align}
    G^{R,A}_\mathbf{k}(\omega) = \left(\mathcal T^\dagger_\mathbf{k}\right)^{-1} \left[ (\omega \pm i\eta) \frac{\tilde I}{2} - \mathcal D_\mathbf{k} \right]^{-1} \mathcal T_\mathbf{k}^{-1}, \\
    G^K_\mathbf{k}(\omega) = -i\eta G^R_\mathbf{k}(\omega) \tilde{\mathcal F}(\omega) G^A_\mathbf{k}(\omega),
\end{align}
\end{subequations}
where $G^K_\mathbf{k}$ is the Keldysh GF. The matrix $\tilde{\mathcal F}(\omega)$ takes the form
\begin{equation}
    \tilde{\mathcal F}(\omega) = \mathcal T_\mathbf{k} \begin{pmatrix}
        \mathcal F_\alpha(\omega) & 0 & 0 & 0 \\
        0 & \mathcal F_\beta(\omega) & 0 & 0 \\
        0 & 0 & \mathcal F_\alpha(-\omega) & 0  \\
        0 & 0 & 0 & \mathcal F_\beta(-\omega)
    \end{pmatrix} \mathcal T^\dagger_\mathbf{k},
\end{equation}
where $\mathcal F_{\alpha,\beta}$ are the distribution functions for either magnon chirality.

The total SKFT action of the AMI including the terms coupling it to the FI layer is then $S_{\rm AMI} = S_{\rm AMI}^{(0)} - S_{\rm AMI}^{(1)}$, where the interaction contribution $S_{\rm AMI}^{(1)}$ is given by
\begin{equation}
    S_{\rm AMI}^{(1)} =  \int\! dt \, \sum_\mathbf{k} \left( v^\dagger_\mathbf{k} \phi_\mathbf{k} + \phi_\mathbf{k}^\dagger v_\mathbf{k} + \sum_\mathbf{q} \phi^\dagger_\mathbf{k} D_\mathbf{k-q} \phi_\mathbf{k} \right). 
\end{equation}
Here, 
\begin{subequations}
\begin{align}
v^\dagger_\mathbf{k} &=\frac{J_I}{2}\sqrt{\frac{S}{2N}}\sum_{n \textrm{ or } m}\left( S^-_n, \, S^+_m, \, S^+_n, \, S^-_m \right) e^{i\mathbf{k \cdot r}_{n,m}}, \\
D_\mathbf{k} &= \frac{J_I}{2N}\sum_{n \textrm{ or } m} \textrm{diag}\Big( -S^z_n, \, S^z_m, \, -S^z_n, \, S^z_m \Big)  e^{-i\mathbf{k \cdot r}_{n,m}} ,
\end{align}
\end{subequations}
where subscript $n$ or $m$ indicate sites in sublattice $A$ or $B$, respectively, and the position vector in the complex exponentials changes sublattice for different vector and matrix elements. After applying the Keldysh rotation to the interlayer interaction term, the AMI magnon fields can be integrated out perturbatively from the Schwinger-Keldysh functional integral to second order in the coupling $J_I$, producing an effective action for the FI
\begin{align}\label{eq:effaction}
    S_{\rm eff} &= S_{\rm FI} + \int\! dt \, \sum_n \mathbf{S}^q_n(t) \cdot \bigg[B^{\rm stag}_n \mathbf{e}_z \nonumber \\
    &+ \int\! dt' \, \sum_{n'} \eta_{nn'}(t,t') \mathbf{S}^c_{n'}(t') - \sum_{n'} \Lambda_{nn'} \partial_t S^{z,c}_{n'}(t)\mathbf{e}_z \bigg] \\
    &+ \frac{i}{2}\int\! dtdt' \, \sum_{nn'} \mathbf{S}^{q}_n(t) C_{nn'}(t,t') \mathbf{S}^{q}_{n'}(t'). \nonumber
\end{align}
The last term on the RHS of Eq.~\eqref{eq:effaction}, encodes thermal and quantum fluctuations of the spins within the FI due to the AMI layer. Such fluctuations are correlated by the tensor
\begin{equation}
    C_{nn'}(t,t') = \begin{pmatrix}
        C^\|_{nn'}(t,t') & C^\perp_{nn'}(t,t') & 0 \\
        -C^\perp_{nn'}(t,t') & C^\|_{nn'}(t,t') & 0 \\
        0 & 0 & C^{zz}_{nn'}\delta(t-t')
    \end{pmatrix}.
\end{equation}
Here, $C^{\|,\perp}_{nn'}(t,t')$ are the nonlocal-in-time fluctuation correlations related to the memory kernel $\eta^\|_{nn'}(t-t')$; $\delta(t-t')$ is the Dirac delta function; and $C^{zz}_{nn'}$ is the time-local fluctuation correlations corresponding to the function $\Lambda_{nn'}$. Since the fluctuation term is quadratic in the quantum fields $\mathbf{S}^q_n$, we neglect it and focus on the deterministic limit.

\subsection{Closed-form solution for nonlocal terms in Eq.~\eqref{eq:llg}}

The staggered magnetic field $B^{\rm stag}_n \mathbf{e}_z$, the memory kernel $\eta_{nn'}(\tau)$, and the nonlocal function $\Lambda_{nn'}$ are given by energy and momentum integrals of different combinations of the AMI magnon GFs. The integrands are sharply peaked in energy space, while the momentum integrals can be simplified by expanding the integrands in powers of $k=|\mathbf{k}|$ and truncating at second order, effectively considering only the effects of long-wavelength AMI magnons. Such simplifications lead to an expression for the staggered field
\begin{equation}
    B^{\rm stag}_n = \pm \frac{J_I}{2\pi \sqrt{v_- v_+}} \int\! dk \, k \left[ \zeta_\pm \mathcal F_\alpha(\tilde\omega) - \zeta_\mp \mathcal F_\beta(\tilde \omega) \right],
\end{equation}
where \mbox{$v_\pm=[1+\mathcal{J}_2(\mathcal K - 2\mathcal J_1 \pm B \delta]/B$} and \mbox{$\zeta_\pm= (B \pm 2\mathcal J_1 \mp \mathcal K )/B$} are dimensionless parameters; $B=\sqrt{\mathcal K(\mathcal{K}-4\mathcal J_1)}$ is half of the magnon gap; and the integral over $\mathbf{k}$ is from $0$ to $\pi/a\sqrt2$. We also get expressions for the non-Markovian kernel
\begin{widetext}
\begin{subequations}\label{eq:kernelint}
\begin{align}
\eta^\|_{nn'}(\tau) &= \begin{dcases}
    \frac{sJ_I^2}{32\sqrt{v_- v_+}} \int\! dk \, k \sin(\tilde\omega_k\tau) \left[ \zeta_\pm J_0(k\tilde r_-) + \zeta_\mp J_0(k\tilde r_+) \right] & \textrm{if $n,n'$ are on the same sublattice}, \\
    \frac{-sJ_I^2}{16 B\sqrt{v_- v_+}} \int\! dk \, k \sin(\tilde\omega_k\tau) \left[J_0(k\tilde r_-) - J_0(k\tilde r_+)\right] & \textrm{if $n,n'$ are on different sublattices},
    \end{dcases} \\
\eta^\perp_{nn'}(\tau) &= \begin{dcases}
    \frac{\pm sJ_I^2}{32\sqrt{v_- v_+}} \int\! dk \, k \cos(\tilde\omega_k\tau) \left[ \zeta_\pm J_0(k\tilde r_-) + \zeta_\mp J_0(k\tilde r_+) \right], & \textrm{if $n,n'$ are on the same sublattice}, \\
    \frac{sJ_I^2}{16 B\sqrt{v_- v_+}} \int\! dk \, k \cos(\tilde\omega_k\tau) \left[J_0(k\tilde r_-) - J_0(k\tilde r_+)\right] & \textrm{if $n,n'$ are on different sublattices}.
    \end{dcases}
\end{align}
\end{subequations}
Here, $\tilde\omega_k = 2B+k^2$; $\mathbf{r}=\mathbf{r}_n-\mathbf{r}_{n'}$ is the relative position vector between sites $n$ and $n'$; $\tilde r_\pm^2 = \mathbf r^T \textrm{diag}(1/2v_\pm, \, 1/2v_\mp) \mathbf r$ is the anisotropically rescaled norm of $\mathbf{r}$; and $J_0(x)$ is the $0$-th Bessel function. In both cases, the upper (lower) signs are for $n$ in sublattice $A$ ($B$). Unlike the memory kernel, the nonlocal function depends on the distribution functions $\mathcal F_{\alpha,\beta}(\omega)$ and is given by
\begin{equation}
\Lambda_{nn'}=\begin{dcases}
\frac{-J_I^2}{64 v_+v_-}\int\! dk \, k \left[ \zeta_\pm^2\partial_\omega \mathcal F_\alpha(\tilde \omega_k) J_0^2(k\tilde r_-) + \zeta_\mp^2\partial_\omega \mathcal F_\beta(\tilde \omega_k) J_0^2(k\tilde r_+) \right] & \textrm{if $n,n'$ are on the same sublattice}, \\ 
\frac{-J_I^2}{16 B^2 v_+v_-}\int\! dk \, k \left[ \partial_\omega \mathcal F_\alpha(\tilde\omega_k)  J_0^2(k\tilde r_-) + \partial_\omega \mathcal F_\beta(\tilde\omega_k) J_0^2(k\tilde r_+)\right] & \textrm{if $n,n'$ are on different sublattices.} 
\end{dcases}
\end{equation}
\end{widetext}

\end{document}

%% file: main.bbl
%